\documentclass[3p,times]{elsarticle}

\usepackage{amssymb}

\usepackage{setspace} 

\usepackage[numbers]{natbib}

\begin{document}

\begin{frontmatter}




\journal{Nuclear Instruments and Methods A}

\title{The Analytical Method algorithm for trigger primitives generation at the LHC Drift Tubes detector}

\cortext[cor1]{Corresponding author}

\author[1] {G.~Abbiendi}
\author[2]{ J.~Alcaraz Maestre}
\author[2]{ A.~\'Alvarez Fern\'andez}
\author[3]{ B.~\'Alvarez Gonz\'alez}
\author[4]{ N.~Amapane}
\author[2]{ I.~Bachiller}
\author[6]{ L.~Barcellan}
\author[1]{ C.~Baldanza}
\author[{1,7}]{ C.~Battilana}
\author[6]{ M.~Bellato}
% ...

\affiliation[1]{organization={INFN Sezione di Bologna, Italy}}
\affiliation[2]{organization={Centro de Investigaciones Energ\'eticas Medioambientales y Tecnol\'ogicas (CIEMAT), Spain}}
\affiliation[3]{organization={Universidad de Oviedo, Instituto Universitario de Ciencias y Tecnolog\'ias Espaciales de Asturias (ICTEA), Spain}}
\affiliation[4]{organization={INFN Sezione di Torino, Italy}}
\affiliation[5]{organization={Universit\`a  di Torino, Italy}}
\affiliation[6]{organization={INFN Sezione di Padova; Universit\`a  di Padova, Italy}}
\affiliation[7]{organization={Universit\`a  di Bologna, Italy}}
\affiliation[8]{organization={Wigner Research Centre for Physics, Hungary}}
\affiliation[9]{organization={Institute of Nuclear Research ATOMKI, Hungary}}
\affiliation[10]{organization={Instituto de F\'isica de Cantabria (IFCA), CSIC-Universidad de Cantabria, Spain}}
\affiliation[11]{organization={Universidad Aut\'onoma de Madrid, Madrid, Spain}}
\affiliation[12]{organization={RWTH Aachen University, III. Physikalisches Institut A, Aachen, Germany}}
\affiliation[13]{organization={Laboratori Nazionali di Legnaro dell'INFN, Italy}}
\affiliation[14]{organization={Institute of Physics, University of Debrecen, Hungary}}
\affiliation[15]{organization={Italian National Agency for New Technologies, Energy and sustainable economic development, Bologna, Italy}}
\affiliation[16]{organization={Now at Boston University, Boston, Massachusetts, USA}}
\affiliation[17]{organization={Now at Centro de Investigaci\'on y de Estudios Avanzados del IPN, Mexico City, Mexico}}
\affiliation[18]{organization={Now at Universit\"at Z\"urich, Zurich, Switzerland}}
\affiliation[19]{organization={Now at Karlsruhe Institute of Technology, Germany}}


\begin{abstract}
The Compact Muon Solenoid (CMS) experiment prepares its Phase-2 upgrade for the high-luminosity era of the LHC operation (HL-LHC). 
Due to the increase of occupancy, trigger latency and rates, the full electronics of the CMS Drift Tube (DT) chambers will need to be replaced. 
In the new design, the time bin for the digitisation of the chamber signals will be of around 1~ns, and the totality of the signals will be forwarded asynchronously to the service cavern at full resolution.
The new backend system will be in charge of building the trigger primitives of each chamber. These trigger primitives contain the information at chamber level about the muon candidates position, direction, and collision time, and are used as input in the L1 CMS trigger.
The added functionalities will improve the robustness of the system against ageing. 
An algorithm based on analytical solutions for reconstructing the DT trigger primitives, called Analytical Method, has been implemented both as a software C++ emulator and in firmware.
Its performance has been estimated using the software emulator with simulated and real data samples, and through hardware implementation tests. Measured efficiencies are 96 to 98\% for all qualities and time and spatial resolutions are close to the ultimate performance of the DT chambers.
A prototype chain of the HL-LHC electronics using the Analytical Method for trigger primitive generation has been installed during Long Shutdown 2 of the LHC and operated in CMS cosmic data taking campaigns in 2020 and 2021. Results from this validation step, the so-called Slice Test, are presented.
\end{abstract}


\end{frontmatter}


\section{Introduction}
\label{sec:introduction}
\label{Intro}
The Compact Muon Solenoid (CMS) is one of two general-purpose detectors at the LHC. 
Muon reconstruction and triggering in CMS are performed with gaseous detectors, installed outside the superconducting 3.8~T solenoid and sandwiched between steel layers. Drift Tube (DT) chambers are used in the central  (or barrel) region characterized by $|\eta|<1.2$. Cathode Strip Chambers (CSC) are located in the endcap region characterized by $0.9 < |\eta| < 2.4 $. Both are complemented by a system of Resistive Plate Chambers (RPCs) covering the range characterized by $|\eta|<1.9$.  
The barrel chambers are arranged in four muon stations (MB1, MB2, MB3, MB4) with increasing radius, embedded in the iron yoke surrounding the superconducting magnet coil. Along the beam axis conventionally labelled as z, DTs and RPCs are divided into 5 slices, called wheels, with Wheel 0  (Wh0) centered at z=0 and wheels Wh+1 and Wh+2 in the positive z direction and Wh-1 and Wh-2 in the negative z direction. Within each wheel, chambers are arranged in 12 sectors in azimuthal angle $\phi$.
The CMS experiment has been designed with a two-level trigger system: the Level-1 Trigger (L1), implemented in custom-designed electronics; and the High-Level Trigger (HLT), a streamlined version of the CMS offline reconstruction software running on a computer farm. 
A more detailed description of the CMS detector can be found in \cite{CMSdes}.
 
\begin{figure}[ht!]
\centering
\includegraphics[width=3.5in]{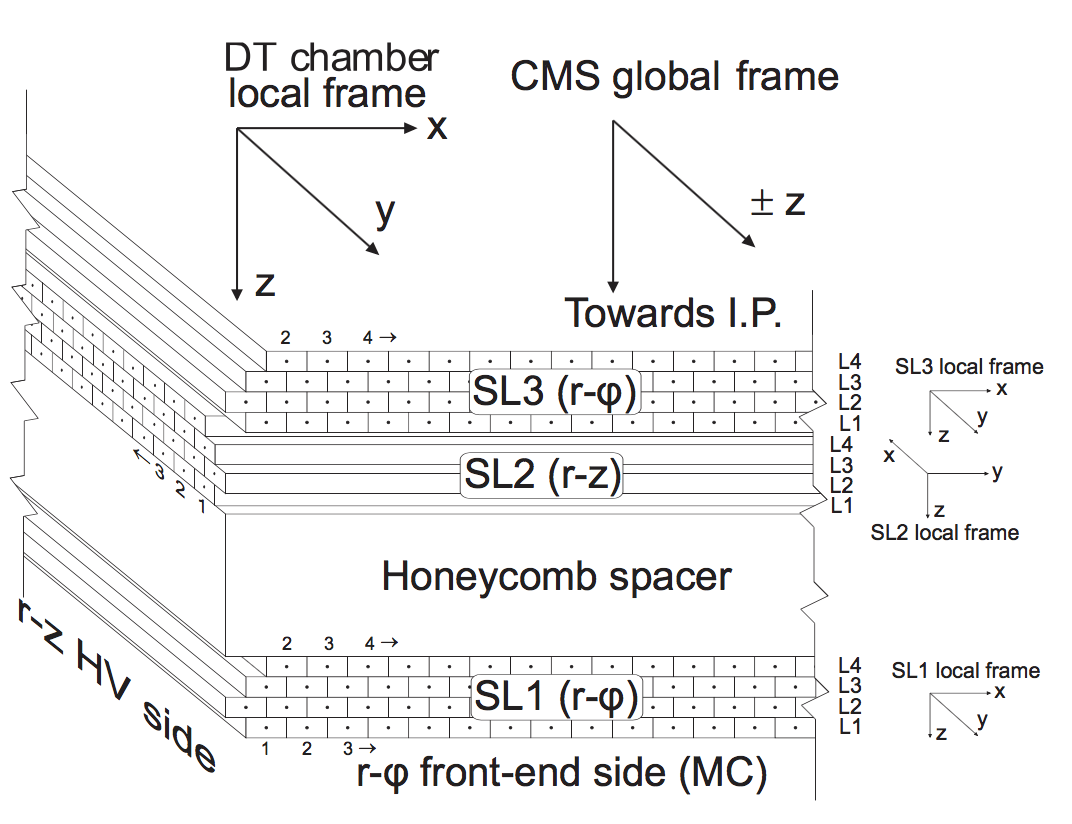}
\caption{Schematic view of a DT chamber.}
\label{Fig:OneChamber}
\end{figure}

The basic element of the DT detector is the drift cell. Its transverse size is $42\times13$ mm$^{2}$ with a 50~$\mu$m diameter gold-plated stainless steel anode wire at the centre, and cathode and strip electrodes glued on the cell walls. The gas is an 85:15 mixture of Ar:CO$_2$, which provides a saturated drift velocity of about 54~$\mu$m/ns. The drift time of electrons produced by ionization is measured by the electronics. With this gas mixture, the maximum drift time is $\sim$390~ns. Hits are then reconstructed, with a left-right ambiguity in their position with respect to the cell wire. Four half-staggered layers of parallel cells form a superlayer (SL), which allows for solving single-hit ambiguities in position by providing the measurement of two-dimensional segments. A chamber is composed by two superlayers measuring the r-$\varphi$ coordinates (SL1, SL3), with the wires parallel to the beamline and separated by $23.5$~cm, and an orthogonal superlayer measuring the r-z coordinates (SL2) in the three inner stations.
A  schematic view of a DT chamber is shown in Fig.~\ref{Fig:OneChamber}

With the High Luminosity (HL-LHC) program, scheduled to start in 2029, the collider will reach unprecedented performance in terms of instantaneous luminosity up to 7.5 $\times 10^{34}$ cm$^{-2}$s$^{-1}$, potentially leading to a total integrated luminosity of up to 4000 fb$^{-1}$ after ten years of operations. 
The Phase-2 upgrade will be a major enhancement of the current muon system, with new Gas Electron Multiplier (GEM) and improved RPC (iRPC) detectors to be installed in the endcap region \cite{MuonTDR, GEMTDR}. One station of GEMs has already been installed in the CMS detector during the Long Shutdown 2 (LS2) of the LHC.
CSC electronics will also be replaced to deal with the increased trigger rates and occupancy in the HL-LHC. For CSCs, the front end upgrade has been completed during LS2, and the backend upgrade will be completed later.
For Phase-2, present DT on detector electronics will be replaced by new boards, the so-called OBDT (On detector Board for Drift Tubes) \cite{OBDTref}, which will perform the time digitisation of the chamber signals inside radiation-tolerant FPGAs with a time bin of around 1~ns. 
All digitised signals will then be asynchronously streamed via high-speed optical links to the backend system, outside the experimental cavern, where the trigger primitives (TP) are generated in commercial FPGAs. 
The readout electronics of the RPC system will also undergo major improvements for the CMS Phase-2, in particular with the increase of the signal sampling frequency, going from 40~MHz to 640~MHz, which will allow to better exploit the detector intrinsic time resolution of 1.5~ns.

Here we present an algorithm, called `Analytical Method' (AM), to perform muon trigger primitive generation for the Phase-2 in the central CMS barrel by using information from the DT and RPC systems. 

Given the improved time digitization of Phase-2 DT input hits of 1~ns, the generated Phase-2 TPs are expected to provide a measurement of the time of the collision generating the muon with a 1 ns granularity, much finer than the 25~ns bins needed for bunch crossing (BX) identification. The muon segment parameters (position and direction) will have a resolution comparable to what is reachable with the present offline reconstruction software~\cite{MuonDPGpaper}.

\section{Description of the Analytical Method algorithm}
\label{sec:AMAlgo}

The input information to the AM algorithm is the wire numbers of the hit cells and the corresponding hit times with respect to the start of the LHC orbit. From these, using the known value of the drift velocity, and assuming a given laterality hypothesis (whether the signal was produced right or left of the cell wire), the hit position is reconstructed.
For a given hypothesis of muon trajectory within a given superlayer, which is a straight segment, the collision time ($t_0$), local segment slope ($\tan \psi$) and local position ($x_0$) can be analytically determined from only three hits in different layers belonging to the muon track.

The algorithm can be logically separated into different steps. 

In the \textbf{Grouping} step, hits in regions of 10 cells in a given superlayer are considered at a time. Hits lying in geometrical patterns consistent with possible muon physical trajectories are selected.

In the  \textbf{Fitting} step muon primitives time $t_0$,  the BX of the corresponding proton-proton interaction that produced the muon, the segment position $x_0$, and the slope with respect to the chamber perpendicular axis $\tan \psi$, are computed in the given superlayer using the least squares minimization method.
For 4 hits candidates, the hit laterality combination resulting in the smallest value for the fitted $\chi^2$ is selected among the possible patterns.
For groups of 3 hits, all hit laterality assumptions providing physical solutions are considered as candidates.

The \textbf{Correlation} step combines the information of the two r-$\varphi$ superlayers. For muon candidates with segments both in SL1 and SL3, segments are combined if the corresponding times are within a window of $\pm$25~ns. 
If a successful correlation is found, a combined trigger primitive is produced. For correlated TPs the parameters are computed as the average of the superlayer segment times and positions, while the direction is obtained as the difference in segment positions in SL3 and SL1 divided by the distance between the two r-$\varphi$ superlayer centers.
The corresponding superlayer input segments are then discarded.
If no match is found, all superlayer TP candidates are kept at this stage.

In the  \textbf{Confirmation} step, remaining superlayer segment candidates that extrapolate to at least 2 hits in the opposite r-$\varphi$ superlayer are tagged.

In addition, cleaning filters are applied at different stages of the algorithm to reduce the amount of fake/duplicated primitives in the output.

At the end of the process, a quality code is assigned as described in the table below.
\begin{center}
\begin{tabular}{ c c c }
 Quality & Description &Type  \\ 
 1 & 3-hit segment & uncorrelated\\
 2 & 3+2 hits segment & confirmed \\
 3 & 4 hit segment & uncorrelated \\ 
 4 & 4+2 hits segment & confirmed \\ 
 5 & non-existing label & \\
 6 & 3+3 hits segment & correlated \\
 7 & 4+3 hits segment & correlated \\
 8 & 4+4 hits segment & correlated  
\end{tabular}
\end{center}

In a final step, information from RPC chambers can be added to define the so-called super-primitives, with time parameter upgraded using the RPC measurement. 
Performance studies for super-primitives have been reported previously~\cite{L1TDR} and are not included here.

The trigger primitive position is finally translated to the CMS global sector coordinates, being $\phi$ the azimuthal angle of the TP position with respect to each sector center, and with the muon bending angle defined as $\phi_B=\psi-\phi$.
The range of the $\phi$ and $\phi_B$ variables is defined by the following scales: 65536 per 0.5 rad for $\phi$ and 4096 per 2 rad for $\phi_B$.
Each trigger primitive consists of 64 bits.
 
\section{Simulation and Monte Carlo samples}
\label{sec:Simulation}

A set of simulated samples was used to estimate the AM performance, for a range of possible Phase-2 conditions, in particular regarding the number of simultaneous proton-proton interactions considered (pile-up). 
For the results presented here, a sample containing 3700 events with four prompt muon pairs per event was used. Each muon pair is formed of 2 back-to-back generated muons, with flat $\phi$, $\eta$, and p$_{\rm T}$  distributions, in the p$_{\rm T}$ range between 2 GeV and 200 GeV and within $|\eta|<1.2$. 
Additional proton-proton interactions are generated within a window of $\pm$16 BX around the central one, fully covering the maximum drift time of ~390 ns, with an average value of 200 pile-up events per bunch crossing. In addition, the GEANT \cite{bibGEANT} simulation configuration takes into account backgrounds from long-lived particles originating from collisions, in particular low-energy neutrons, that can produce hits in the DT chambers that accumulate uniformely distributed in time over the LHC orbit~\cite{MuonDPGpaper}.
The simulation assumes a perfect inter-chamber calibration. 

DT offline segments as computed using the standard DT reconstruction algorithm were taken as reference for measurements of the trigger primitive performance.

\begin{figure}[ht!]
\begin{center}
\includegraphics[width=3.5in]{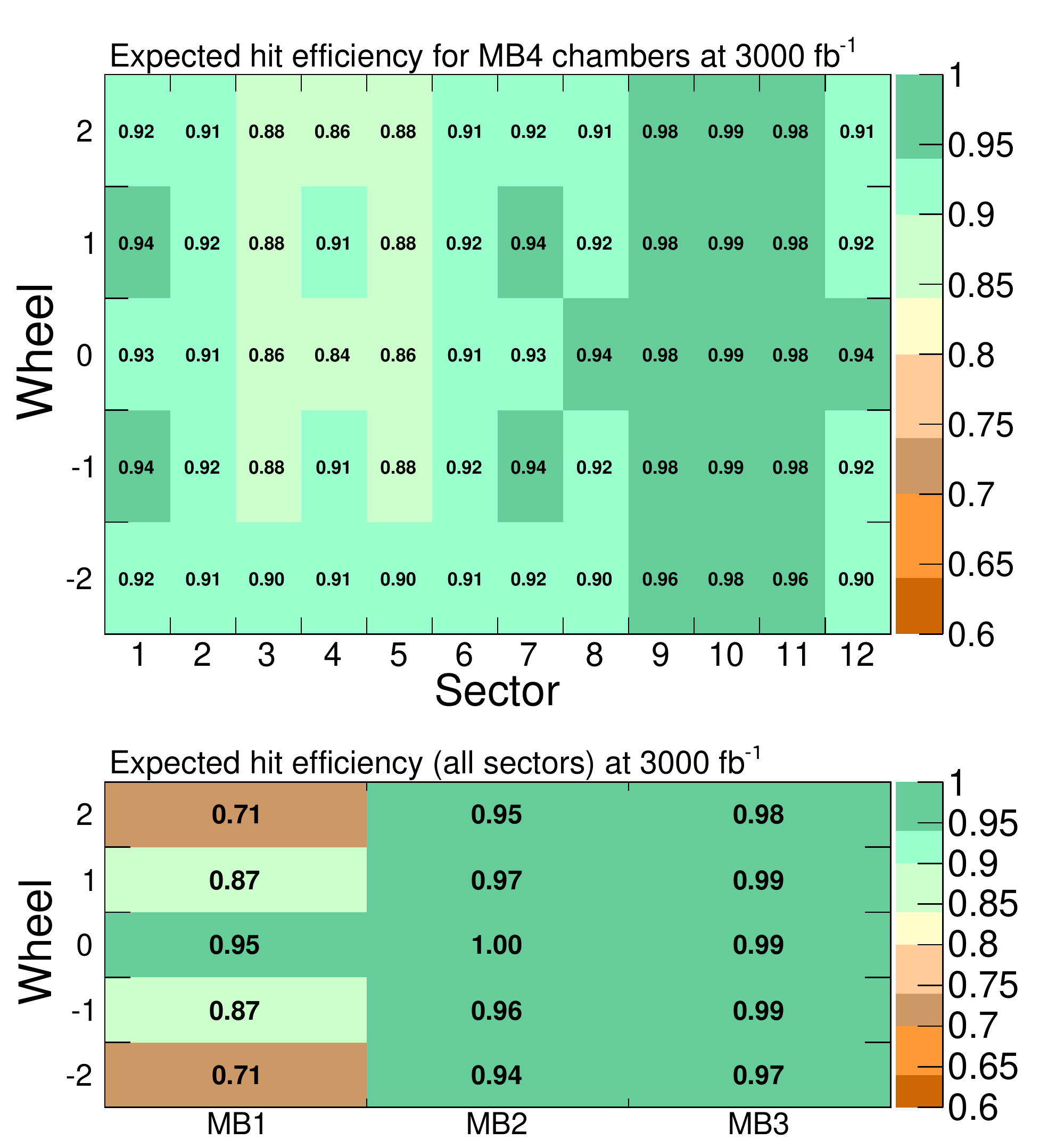}
\end{center}
\caption{Expected hit efficiencies at the end of the HL-LHC for all the DT chambers in CMS. The upper plot shows MB4 chambers, the lower shows MB1, MB2 and MB3, for the conservative ageing scenario described in the text.}
\label{fig_ageing}
\end{figure}

During HL-LHC operation DT chambers will be exposed to high levels of radiation. The accumulated charge from this radiation can result in significant ageing effects that degrade the DT cell performance and lower the DT chamber efficiency. Measurements of the DT detector performance under high radiation have been conducted in the new CERN Gamma Irradiation Facility (GIF++) \cite{GIFpaper} and \cite{GIFpaper2}.
According to these results, ageing scenarios can be simulated by randomly removing DT hits, according to probabilities obtained by the expected inefficiencies in each detector region. These inefficiencies depend on the expected accumulated charge as measured at GIF++. The scenario considered here for TP generation studies corresponds to extreme ageing effects in the DT detector which might arise at the end of the HL-LHC operation.
In this scenario, the lowest DT chamber efficiencies are of the order of 70\% in station 1 (MB1) of the external DT wheels (Wh$\pm$2), increasing to 90\% in some sectors of station 4 (MB4) and remaining significantly higher for the rest of the DT chambers, as can be seen in Fig.~\ref{fig_ageing}.
This is a conservative scenario targetting worst case limits and in which the uncertainties are well contained within the safety factors that have been applied. These hit efficiencies were estimated considering a safety factor of 2 for the expected HL-LHC instantaneous luminosity (2*5 $\times 10^{34}$ cm$^{-2}$s$^{-1}$) and a safety factor of 2 for the expected integrated luminosity (2*3000 fb$^{-1}$).

Based on this data, a degradation of the hit detection efficiency is only expected, by the end of HL-LHC, in the most exposed CMS DT chambers. However, thanks to the redundancy of the system and to the mitigation measures currently implemented as described in section 3.2.4 of~\cite{MuonTDR}, a good performance of the muon triggering and reconstruction is still expected during the whole HL-LHC phase.

\subsection{Efficiencies to trigger on prompt muons. }
\label{sec:EffSingMu}

In order to compute TP efficiencies, the denominator is defined as the number of DT offline segments with at least 4 hits in the r-$\varphi$ view and also 4 hits in the  r-z view (if present) that are geometrically matched with a generated muon with p$_\mathrm{T}> $20~GeV, within a  window of 0.15 in $\eta$ and a window of 0.1 rad in $\phi$. 
In order to select a clean sample and to further suppress bad quality segments arising from out-of-time pile-up events, a cut on the reconstructed segment time of $\pm$15~ns was applied. 
The numerator of the efficiency is defined as the number of trigger primitives with fitted time at the collision BX and matching these segments within a $\phi$ window of 0.1 rad.

\begin{figure}[ht!]
\begin{center}
\includegraphics[width=3.5in]{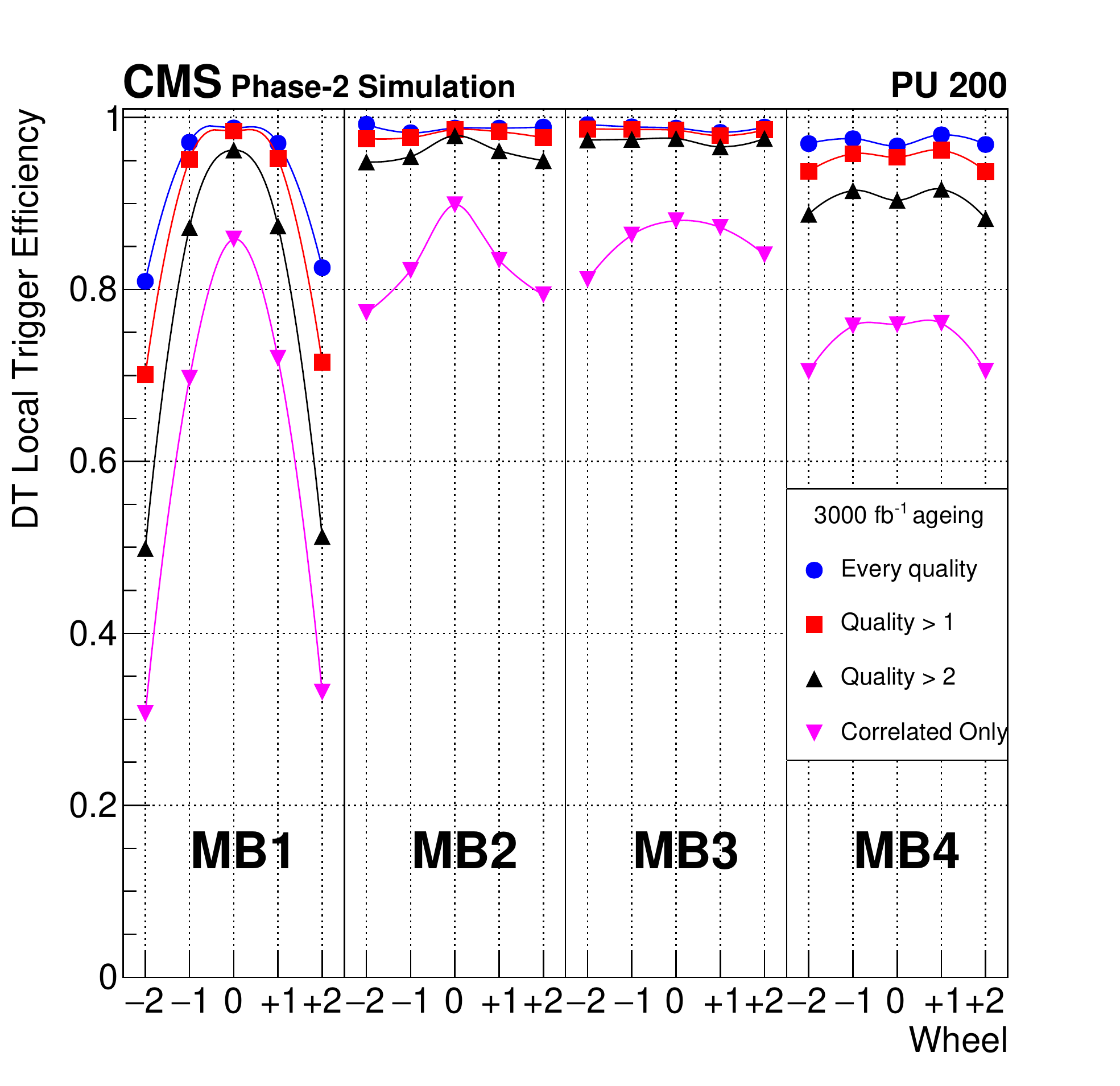}
\end{center}
\caption{ TP efficiency with respect to offline segments for an ageing scenario corresponding to 3000 fb$^{-1}$, for different thresholds on the TP quality. Statistical errors uncertainties are smaller than the markers size. Systematic errors depend mainly on the applied aging uncertainty which, being large, is estimated assuming a worst case limit as described in the text.}
\label{singleMuoneff_qualities}
\end{figure}

When no ageing is considered TP efficiencies are 98$\%$ or higher for a TP quality threshold of 2 or less, while the efficiency for correlated TPs (Quality$>$6) is above 80$\%$ in the whole detector.
TP efficiencies are expected to decrease with ageing.
Fig.~\ref{singleMuoneff_qualities} summarises the DT Phase-2 TP efficiency per station and wheel for different TP quality thresholds for the ageing scenario corresponding to 3000 fb$^{-1}$.  As expected, the efficiency drop is larger in the chambers more affected by ageing effects, in particular for high TP quality thresholds. The efficiency drop can be substantially recovered by accepting confirmed primitives (Quality$>$1).
In addition, previous studies based on Phase-1 TPs~\cite{L1TDR} show that the net effect of the ageing on L1 Muon trigger efficiency is small, dropping $< $5\% in the barrel-endcap overlap region, and less elsewhere.

\subsection{Comparisons with offline segments and intrinsic resolution studies.}
\label{sec:Res}

\begin{figure}[ht!]
\begin{center}
\includegraphics[width=0.35\textwidth]{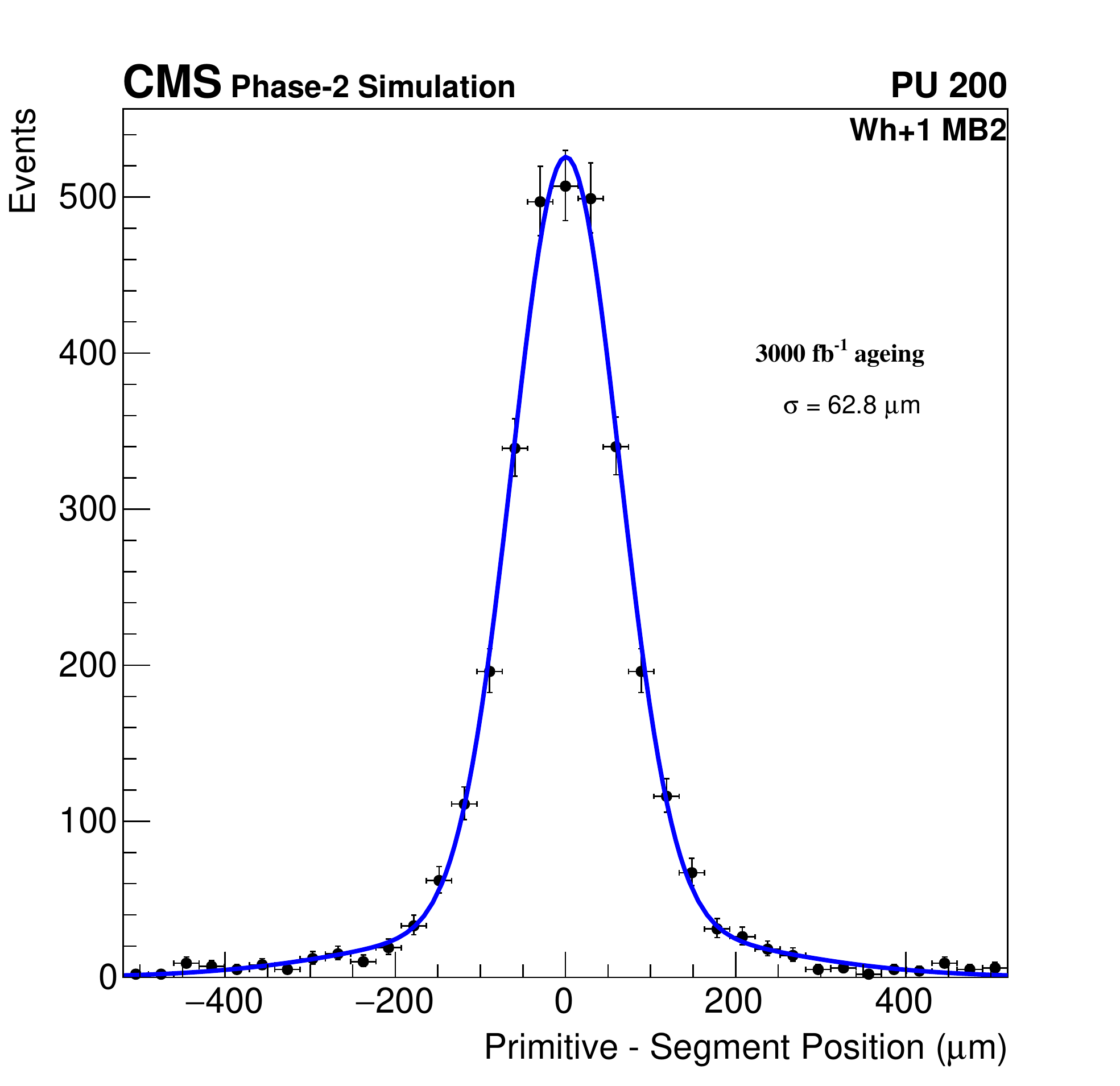}
\includegraphics[width=0.35\textwidth]{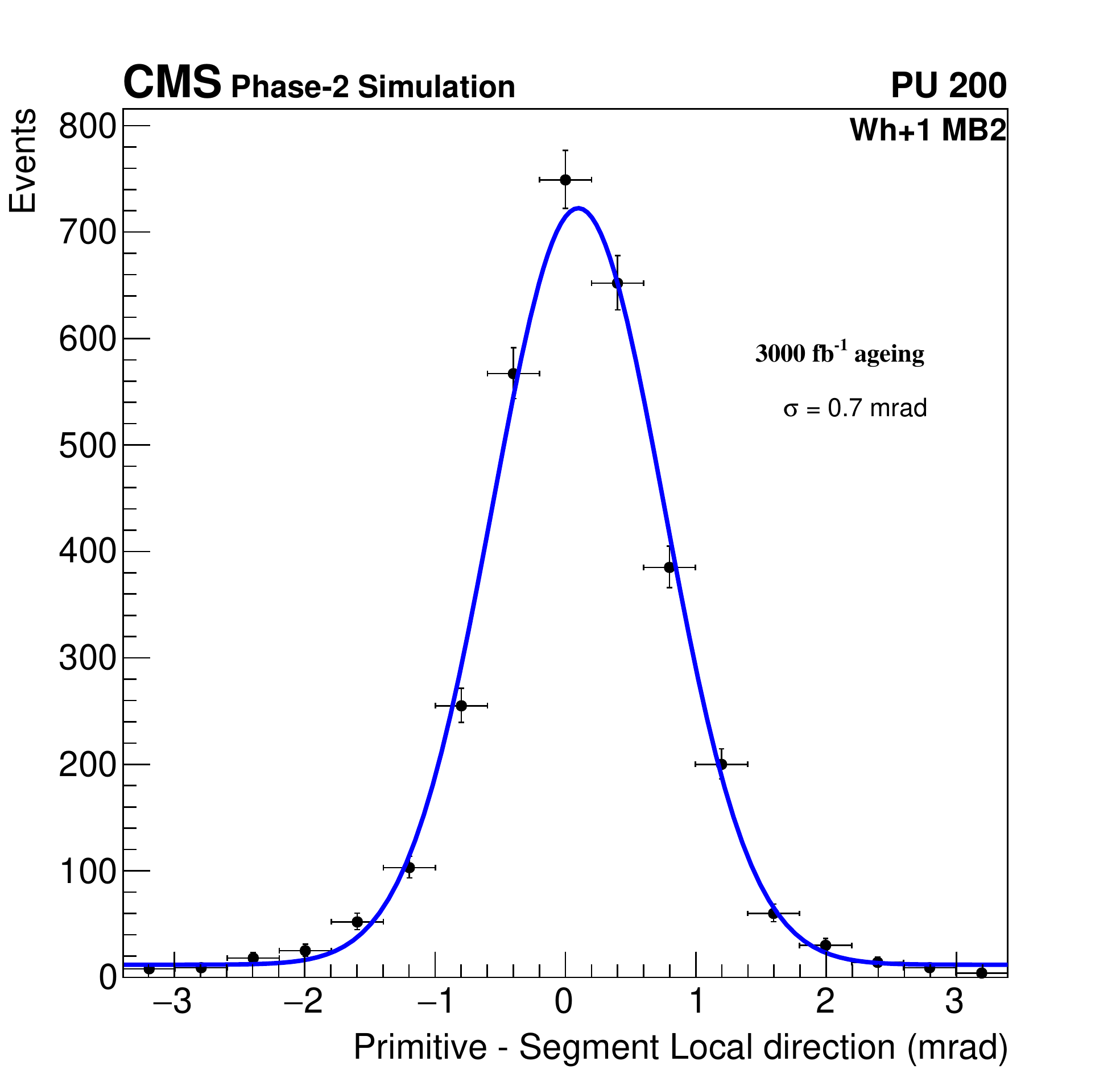}
\end{center}
\caption{Difference with respect to offline reconstructed segments of TP local position $x_0$ (upper) and TP local slope  $\tan \psi$ (lower), for correlated TPs in Wh+1 MB2. End of HL-LHC ageing is applied to hits before TP generation.}
\label{Gauss_resol_posslope}
\end{figure}

To evaluate the quality of the TP fit results, the differences in TP position and angle with respect to the corresponding values of offline reconstructed segments are computed. For this purpose, offline segments are selected satisfying the same matching requirements and quality criteria as for efficiency studies.
As an example, Fig.~\ref{Gauss_resol_posslope}  shows the difference in local position $x_0$ (lower panel) and local slope $\tan \psi$ (upper panel) for correlated TPs in Wh+1 MB2. 
Ageing effects have been applied to hits before the TP generation. Hits have been removed according to the probability obtained by the inefficiency expected in that detector region at the end of HL-LHC. This inefficiency depends on the accumulated integrated charge as measured at GIF++ and quoted in Fig.~\ref{fig_ageing}.
The standard deviation ($\sigma$) of the Gaussian fit to the local position difference distribution is 63~$\mu$m. The $\sigma$ of the Gaussian fit to the local slope difference distribution is 0.7~mrad.

\begin{figure}[ht!]
\begin{center}
\includegraphics[width=0.35\textwidth]{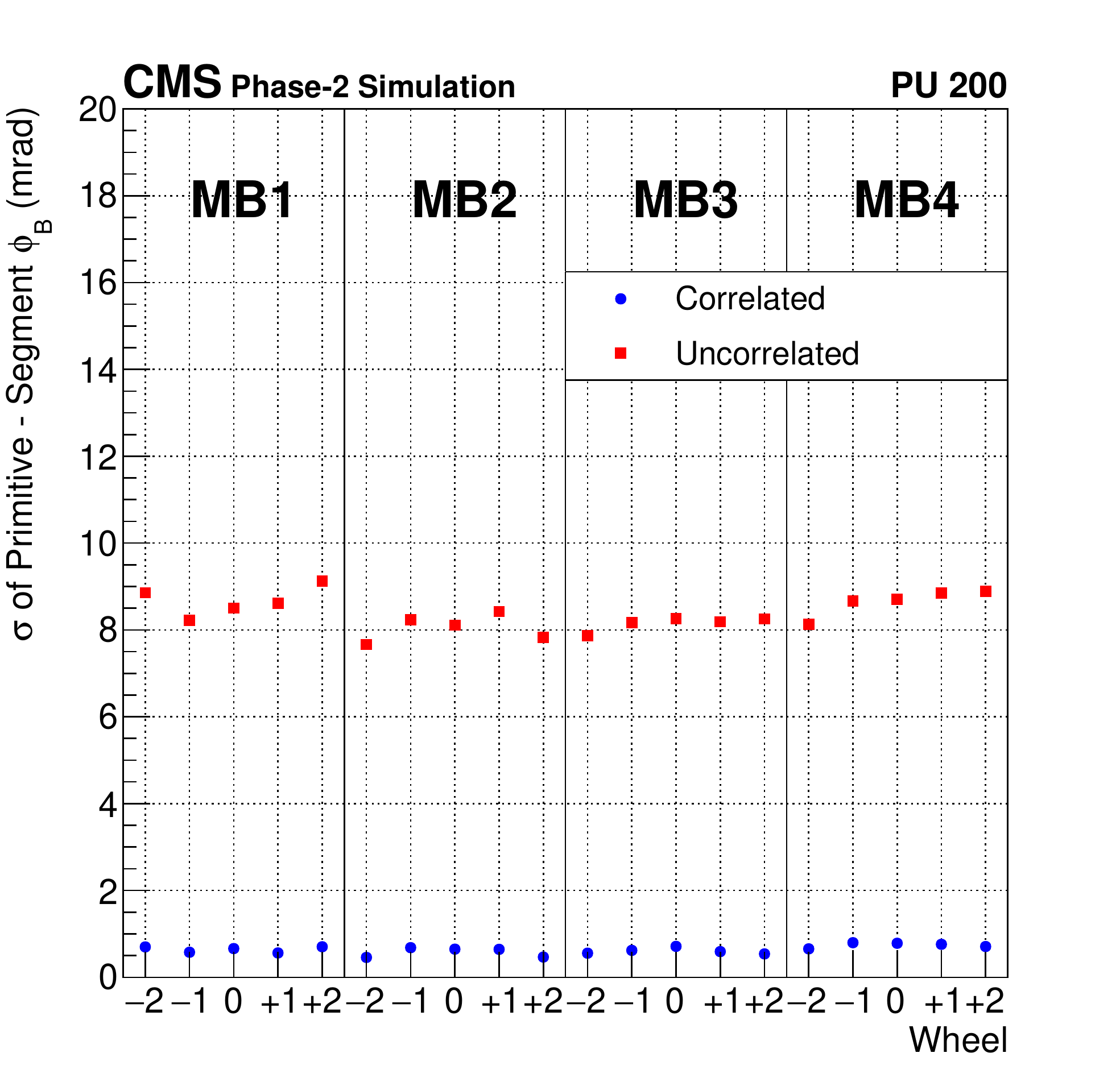}\qquad
\end{center}
\caption{$\sigma$ of the Gaussian fit of the offline segment to TP difference distribution for the bending angle $\phi_B$ for correlated TPs in blue and non-correlated TPs in red. End of HL-LHC ageing is applied to hits before TP generation.}
\label{Gauss_resol_dir}
\end{figure}

Fig.~\ref{Gauss_resol_dir} shows the $\sigma$ of the Gaussian fit to the bending angle $\phi_B$ difference distribution for chambers in the same wheel and station. 
Correlated TPs are shown in blue and non-correlated TPs in red. The better measurement obtained with correlated TPs is due to the effect of the larger level arm between SL1 and SL3 with respect to the measurement in a single superlayer.

In general, the results for position and angular parameters show no significant degradation for aged chambers.

\begin{figure}[ht!]
\begin{center}
\includegraphics[width=0.35\textwidth]{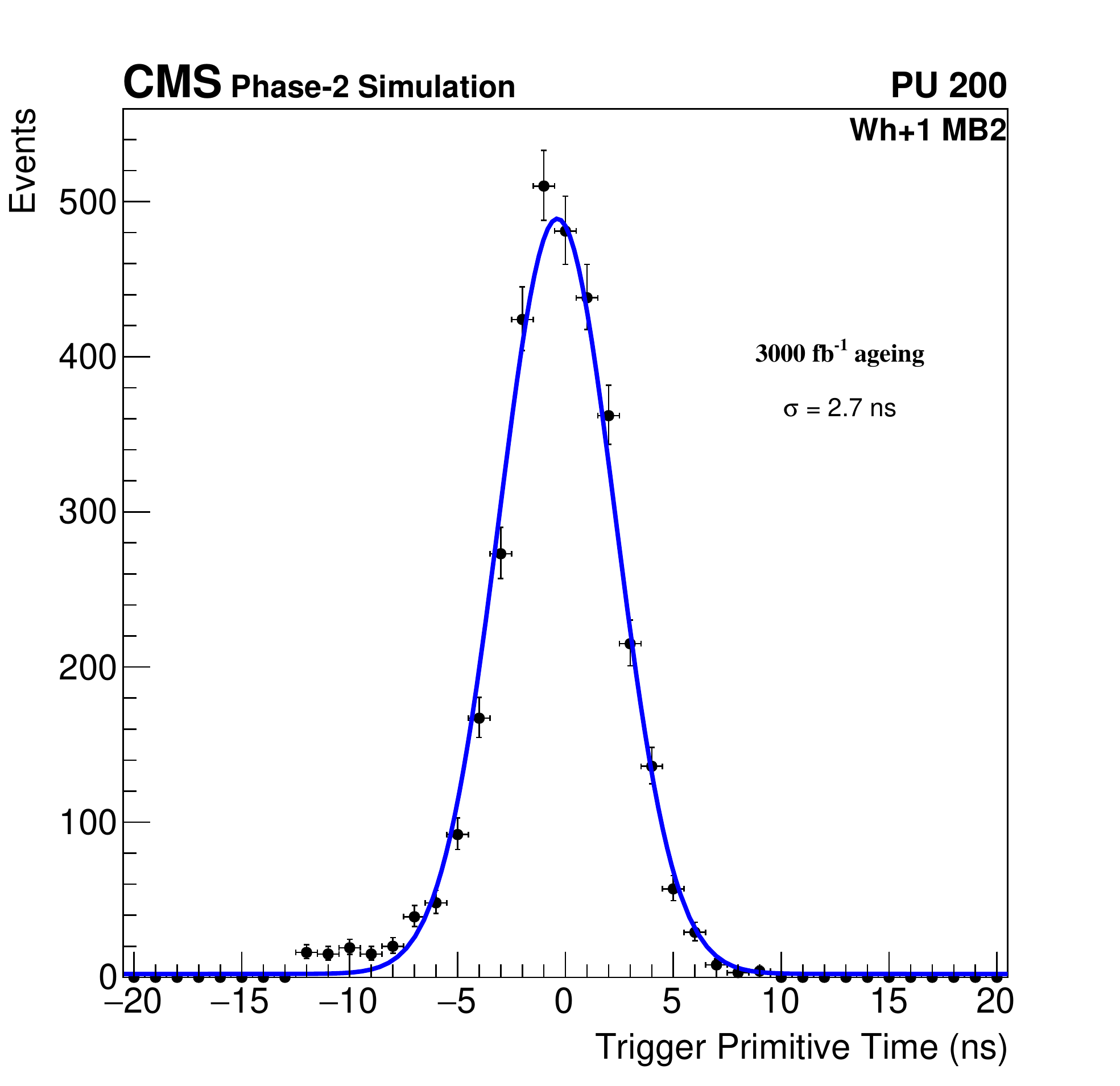}\qquad
\end{center}
\caption{TP $t_0$ distribution for correlated TPs in Wh+1 MB2. End of HL-LHC ageing is applied to hits before TP generation.}
\label{Time}
\end{figure}

Fig.~\ref{Time} shows the time $t_0$ distribution of correlated TPs. The $\sigma$ of the Gaussian fit is 2.7~ns.

It must be stressed that TP to segment differences are not to be considered as intrinsic measurements of the TP parameter resolution, as significant correlations between TPs and the offline segments are to be expected due to sharing of hits. 

\begin{figure}[ht!]
\begin{center}
\includegraphics[width=3.5in]{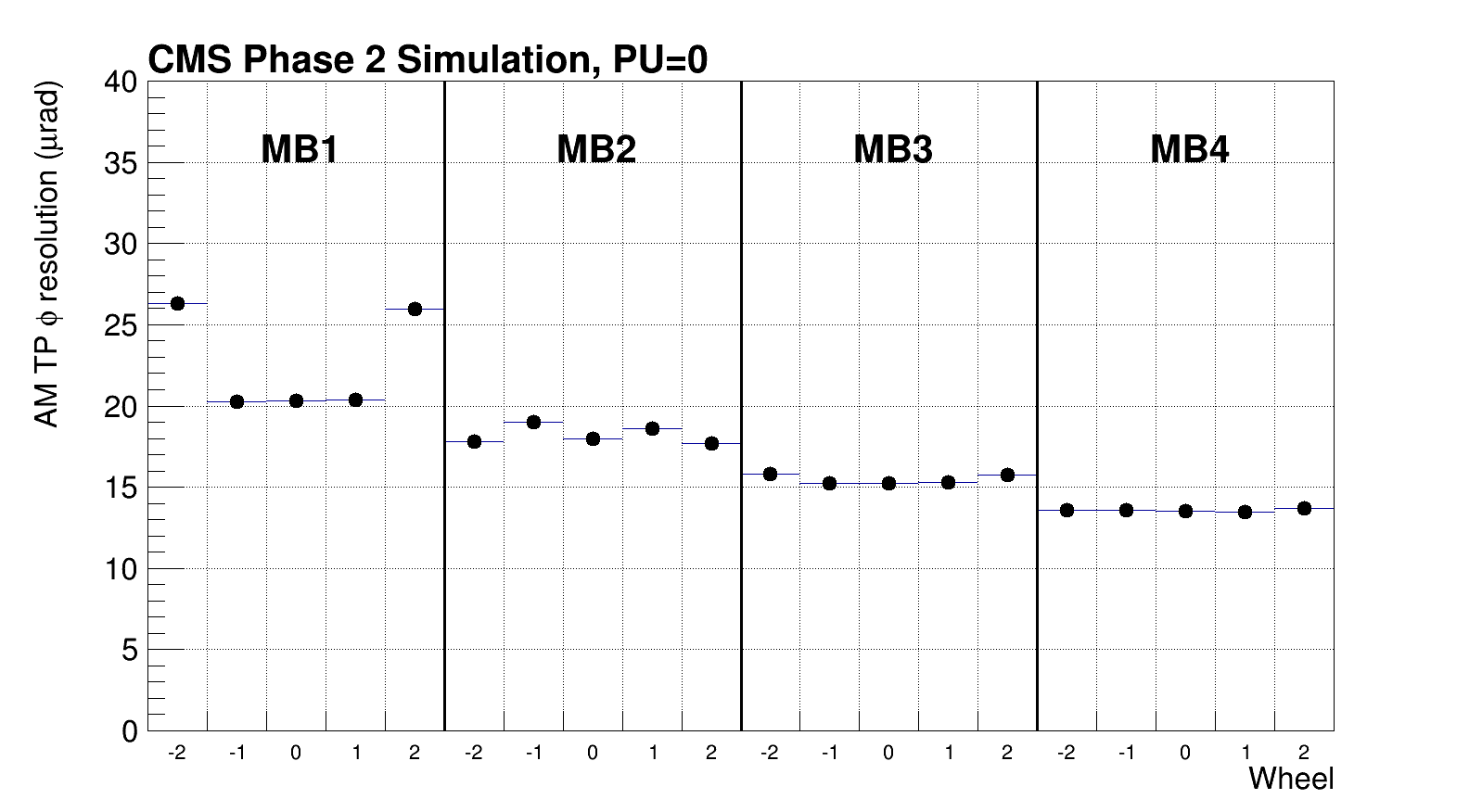}
\includegraphics[width=3.5in]{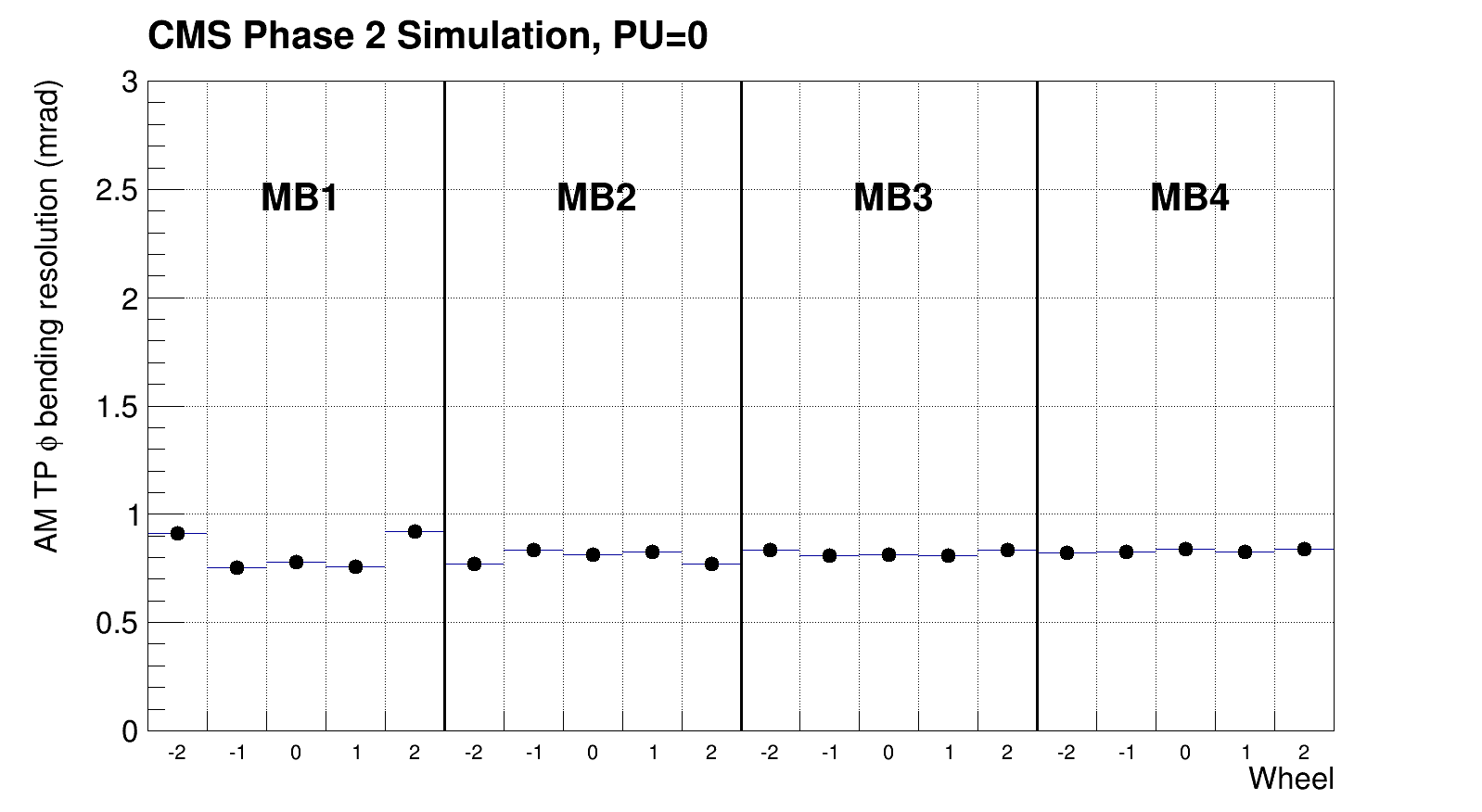}\qquad
\end{center}
\caption{TP sector $\phi$ (upper) and bending angle $\phi_B$ (lower) resolutions for all TP qualities, obtained as the $\sigma$ of the Gaussian fits to the TP to `true muon trajectory' residuals. Ageing effects are not considered in the TP formation.}
\label{fig_resol}
\end{figure}

Dedicated studies using as reference simulated quantities were performed as well. The sample used consisted of 200,000 events containing two muon pairs with flat $\phi$, $\eta$, and p$_{\rm T}$  distributions, and without any contribution from pile-up events. The positions of simulated muon hits in the inner and outermost layers of the DT chambers determine a straight line that is considered as the `true' muon trajectory, and taken as a reference. Fig~\ref{fig_resol} shows the $\sigma$ of the Gaussian fit to the residual of the TP sector $\phi$ (upper) and  $\phi_B$ (lower) distributions with respect to this reference. 
Sector $\phi$ resolution is $<$ 50 $\mu$rad and a $\phi_B$ resolution $<$1~mrad when considering all TP qualities in all stations and wheels. The observed trend on $\phi$ resolution is primarily given by the increasing radial position of the stations from MB1 to MB4. For a given local position resolution in a chamber, the radial position of the station determines the angular variable resolution. The resolutions results presented show the algorithm performance in ideal conditions, without aging, which are comparable to those obtained offline and beyond what is needed for a trigger system. It is to be noted that the aging model applied here is not expected to model perfectly the resolution degradation as it does not include the degradation of the linearity between time and position inside the drift cell.

\subsection{Rate studies}
\label{sec:Rate}

Trigger primitive rates were calculated for the AM algorithm, considering and not considering ageing effects. The same simulated sample used for the efficiency studies described above, was used for rate evaluation. For each event, only chambers not crossed by offline reconstructed muons were considered.
When ageing effects are not considered, the rates for DT TPs with at least 4 hits are in fair agreement with the ones extrapolated from Phase-1 data, of the order of $\sim$0.5~MHz at 10 $\times 10^{34}$ cm$^{-2}$s$^{-1}$ in MB1 external wheels~\cite{MuonTDR}.  
Rates elsewhere are at least a factor of 2 smaller, with rates for most of the chambers being one order of magnitude below the MB1 external wheels.
Ageing reduces the trigger primitive rates, in particular in regions more affected by radiation effects, as expected by the hit efficiency loss. For the MB1 of the external wheels the estimated rate reduction is about factor 3.
All estimated rates are within specifications for the L1 Phase-2 trigger system~\cite{L1TDR}.

\section{Firmware implementation}
\label{sec:Fw}

The Drift Tube AM trigger algorithm, already designed with a firmware-oriented approach, was ported to VHDL in order to estimate the FPGA resources required by the new backend electronic system.
This implementation was validated in prototype boards with real data using emulator-to-firmware comparisons.

The present fimware implementation performs the generation of a TP in the r-$\varphi$ view of the DT chamber and is consistent with the algorithm description provided in section~\ref{sec:AMAlgo}. Some small differences are present, mainly due to practical constraints. In particular, confirmed qualities, r-z view, and super-primitive building out of the RPC information are not implemented in the firmware at the moment.

The first implementation of this algorithm was performed in a Virtex 7 FPGA, in particular, the XC7VX330T-3FFG1761E. TM7~\cite{TM7Triossi} backend modules that were designed for the CMS Phase-1 upgrade were used as a test bench for the real HL-LHC system. 
The TM7 backend modules programmed with the AM trigger algorithm are called AB7 (AM algorithm Beta on Virtex 7).
On top of the AM algorithm logic, several functionalities were included to allow the control and operation of the system.

A series of studies were performed at a test stand at CIEMAT to assess the level of agreement between the current emulator and AB7 firmware implementations of the AM algorithm. 
The input is given by the  DT hits from 10,000 reconstructed $Z\to \mu\mu$ events in the 2016 collision data sample, coming  from all chambers in the CMS detector.
These hits are stored in a file and injected at the input buffers of the AB7 board, modifying the corresponding parameters for each chamber accordingly. The hits have Phase-2 data format and are injected at a predefined time, emulating OBDT behaviour. Hits go through all the chamber trigger chain inside the Virtex 7 FPGA, and trigger primitives are generated in the board and subsequently readout.
The emulator is also run on the same sample.

To compare firmware and emulator,  studies have currently been restricted to events with $<$12 hits in the considered chamber, in order to assess first possible logical differences in firmware and emulator implementations and avoid, at this stage, any effect related to late arriving hits or buffer overflows.
For any given primitive output by the emulator, a corresponding primitive is searched for in the firmware output, sharing the same fitted hits with the same laterality assignments.
The percentage of emulator-firmware primitive pairs matched in this way is greater than 97$\%$ for all qualities. 

\begin{figure}[ht!]
\begin{center}
\includegraphics[width=0.35\textwidth]{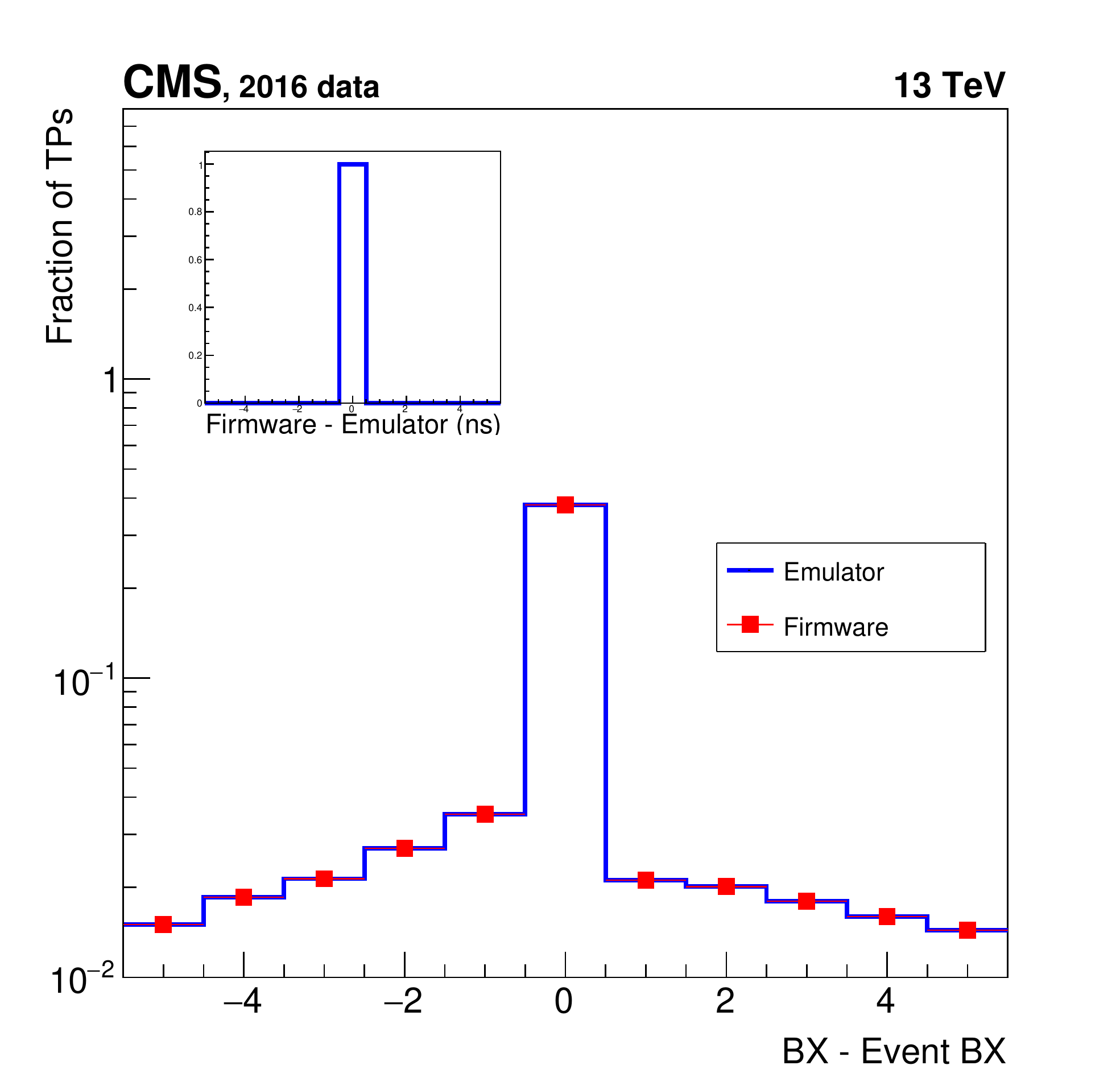} \qquad
\includegraphics[width=0.35\textwidth]{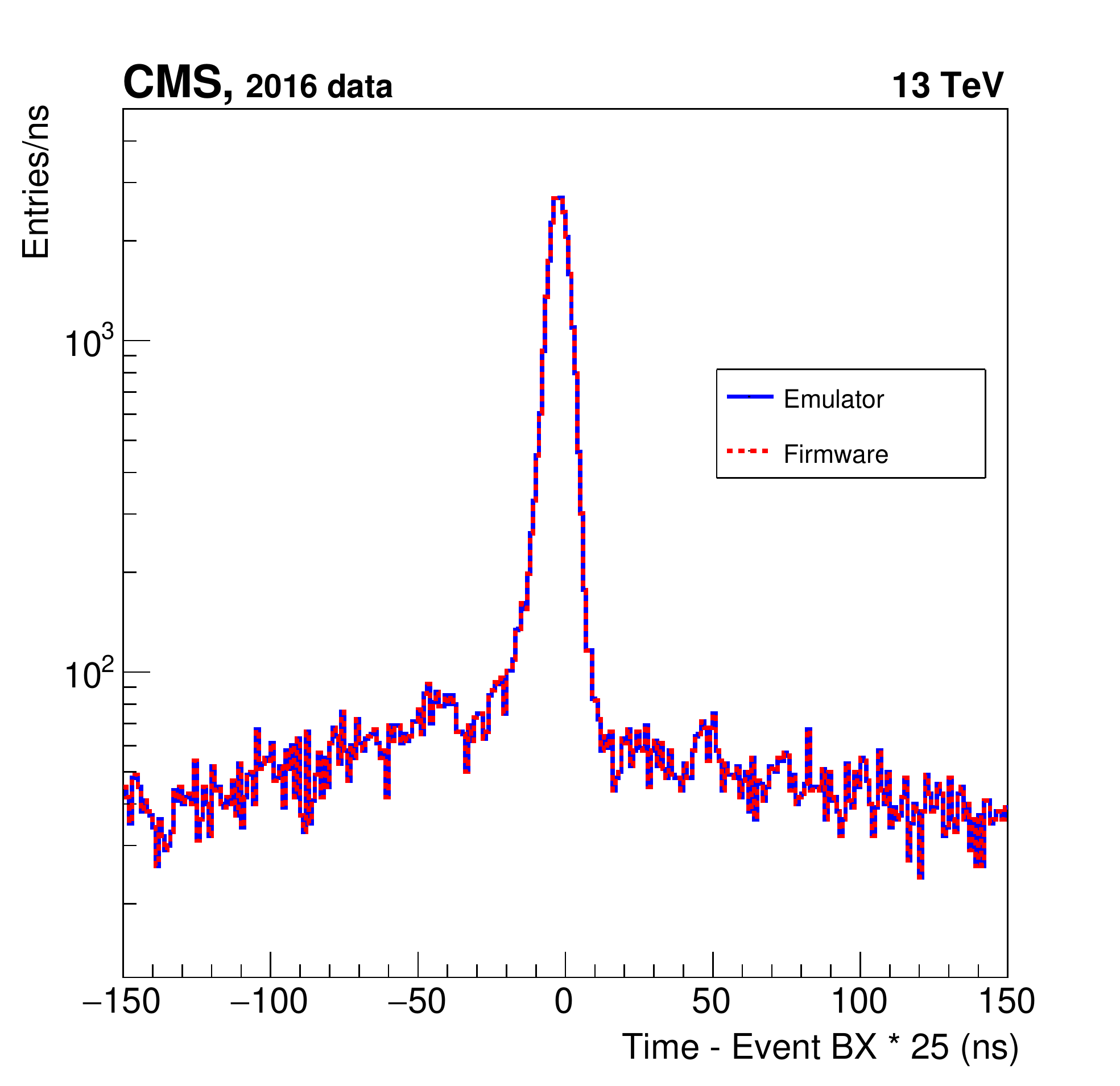} 
\end{center}
\caption{Left: Difference in BX assignment between emulator primitives and event BX (blue) and firmware primitives and event BX (red). Insert shows agreement in the fitted time value at the level of Least Significant Bit (1~ns). Right: trigger primitives time minus Event BX*25 (ns), as obtained by the software emulator (blue) and obtained by the firmware (dashed red). In both plots, only pairs of primitives fitting the same hits with the same laterality (67066 entries) are considered. }
\label{fig:BXtimefwemu}
\end{figure}

In the left panel of Fig.~\ref{fig:BXtimefwemu}, the blue curve shows the difference between trigger primitives BX as obtained by the emulator and the event BX. Red points show the difference between trigger primitives BX obtained by the firmware and the event BX.
As shown in the insert, agreement in time is at the level of the least significant bit (1~ns) for all trigger primitive qualities. 
The blue curve on the right panel shows trigger primitives time as obtained by the software emulator after subtraction of the BX number associated with the event and multiplied by 25~ns (in order to convert from BX number to units of ns). The dashed red curve shows the same quantity for TPs  as obtained by the firmware. Perfect agreement can also be seen.

\begin{figure}[ht!]
\begin{center}
\includegraphics[width=0.35\textwidth]{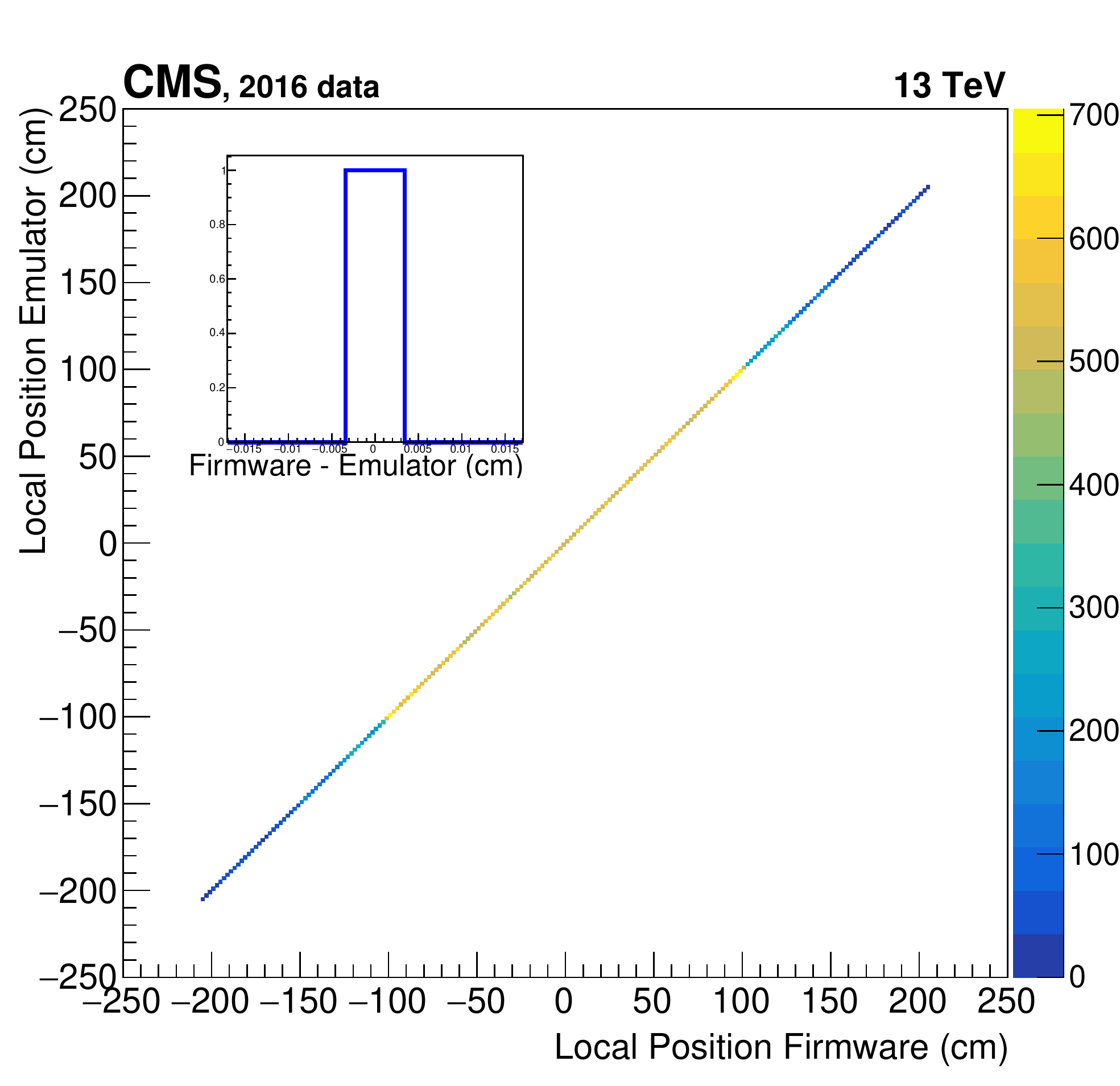} \qquad
\includegraphics[width=0.35\textwidth]{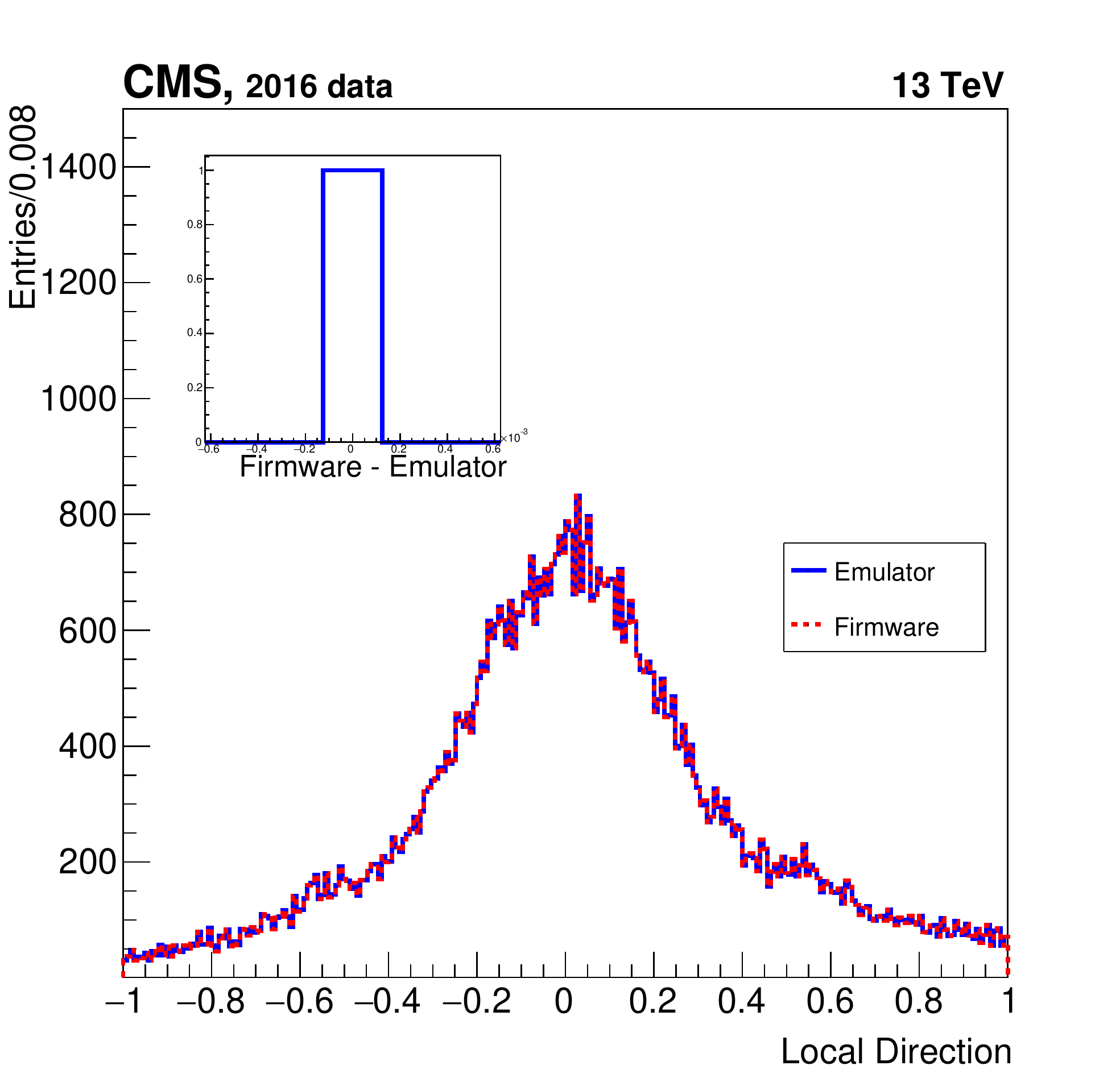} \\
\end{center}
\caption{Left: Primitives local position $x_0$ as obtained by the software emulator versus trigger primitives local position obtained by the firmware. Right: Trigger primitives local direction $\tan \psi$ as obtained by the software emulator (blue) and obtained by the firmware (dashed red). Inserts show agreement at the level of the Least Significant Bit for all trigger primitive qualities for both variables. In both plots only pairs of primitives fitting same hits with the same laterality (67066 entries) are considered. }
\label{fig:PosSlopefwemu}
\end{figure}

In Fig.~\ref{fig:PosSlopefwemu}, the left panel shows trigger primitives local position as obtained by the software emulator versus trigger primitives local position as obtained by the firmware. The right panel shows trigger primitives local direction as obtained by the software emulator (blue) and obtained by the firmware (dashed red). Again the corresponding inserts show agreement at the level of the Least Significant Bit for all trigger primitive qualities for both variables.

\section{Operation of the AM trigger in the CMS DT Slice Test}
\label{sec:ST} 
During the Long Shutdown 2 (LS2) of the LHC, a complete exercise was made to instrument one sector (Wh+2, S12) of the CMS detector with the Phase-2 DT electronics prototypes. 
Chamber signals were split into the four DT stations, so they can be readout at the same time with the legacy electronics and with the new OBDT boards.
In the current setup, each OBDT covers one superlayer, except in the MB4 station, which required two OBDTs per SL.
The time digitization performed by the OBDT has a time bin of $\sim$0.8~ns (25/30~ns) and is referenced to the Bunch Crossing 0  signal distributed by the CMS clock and control system (TCDS) at the LHC orbit frequency of 11.22 kHz.
The OBDT outputs the timestamps of the detected  hits, that are sent via high-speed optical links through a GBTx protocol to the backend system.

Several prototypes of the backend boards based on the TM7 boards were used either for Slow Control, so-called MOCO (Monitoring and Control) board, or for readout and triggering purposes through AB7 boards. 
Five AB7 boards readout the full sector, one for MB1, one for MB2, one for MB3, receiving inputs from 3 OBTDs each, and two for the MB4, each of them reading two OBDTs.
The AB7 board is in charge of performing the trigger primitive generation first at the level of SL by reading hits from each OBDT, and then by combining two of them to obtain correlated TP candidates from the two r-$\varphi$ superlayers.

This prototype system was integrated into the trigger and data acquisition (DAQ) system of CMS. Cosmic data-taking campaigns were performed regularly during several months in LS2. 
In normal operation mode, the trigger decision was defined by different conditions provided by the legacy system electronics.
In parallel the TPs obtained by the Phase-2 system were computed and sent to the CMS DAQ using the AB7 board.

\begin{figure}[ht!]
\begin{center}
\includegraphics[width=0.35\textwidth]{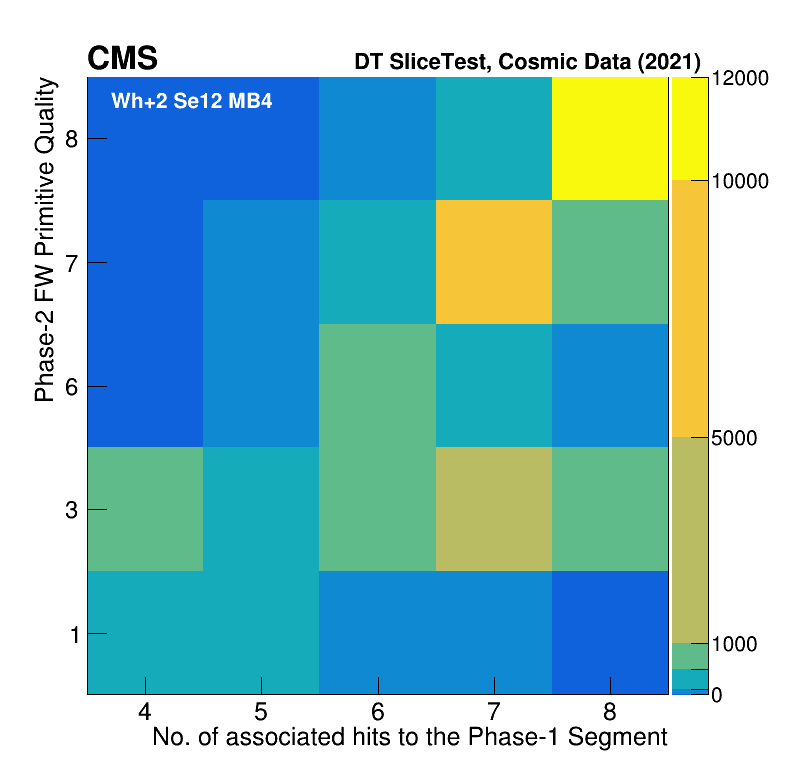} \qquad
\includegraphics[width=0.35\textwidth]{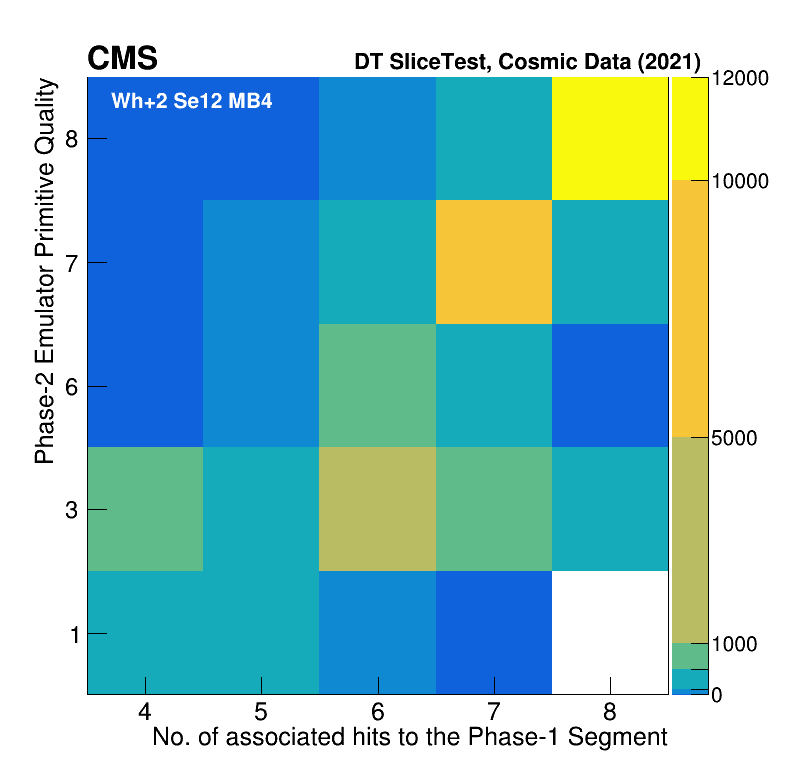} \\
\end{center}
\caption{2D distribution of the Phase-2 TP quality obtained by the AM 
firmware (upper) and emulator (lower) versus the number of hits associated with the offline reconstructed segment in the r-$\varphi$ view of the MB4 station for the DT Slice Test data taken in 2021}
\label{fig:STTPQualNhitsSeg}
\end{figure}

Fig.~\ref{fig:STTPQualNhitsSeg} shows the 2D distribution of the Phase-2 TP qualities, as obtained by the AM algorithm in MB4, as a function of the number of hits associated with the offline reconstructed segment.  Offline segments must have at least 4 hits in the r-$\varphi$ view. Only the segment with more associated hits, and the TP with the highest quality are considered.
In the upper panel, TPs were obtained by the firmware running on the AB7, while on the lower panel TPs were obtained by running the emulator code on the hits sent by the corresponding OBDT boards.
In both cases, a good correlation between TP qualities and the number of this in the offline segment is observed.

\begin{figure}[ht!]
\begin{center}
\includegraphics[width=3.5in]{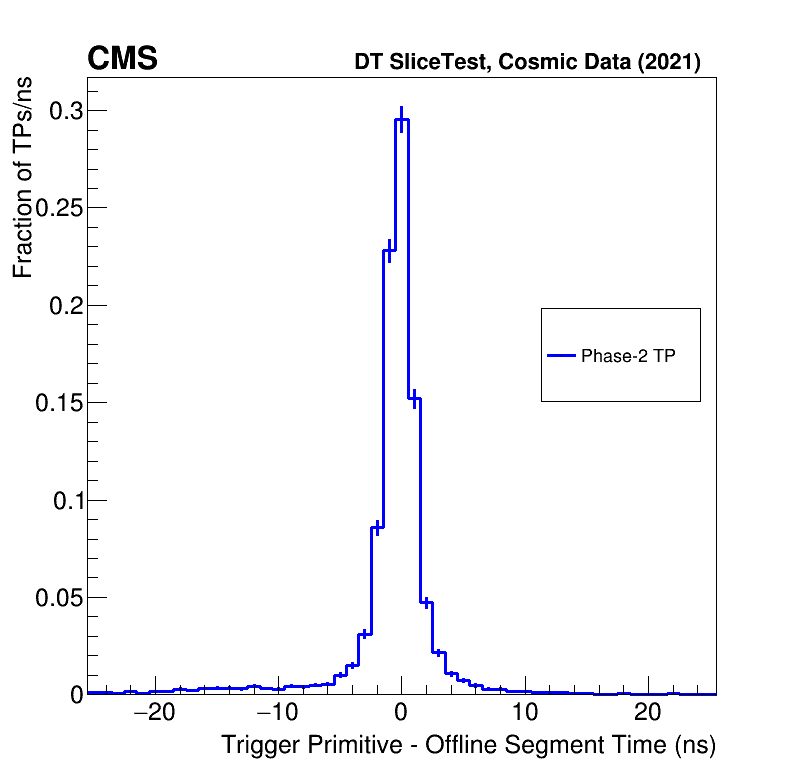} \qquad
\end{center}
\caption{Difference between the AM TP fitted time $t_0$ and the offline reconstructed segment time for the MB4 station in the Slice Test setup during 2021 cosmics data taking. Larger tail at negative values is compatible with the known effect of delta rays and detector deadtime.}
\label{fig:STTime}
\end{figure}

An example of the timing performance of the system can be seen in Fig.~\ref{fig:STTime}, in which the difference between the AM TP fitted time and the offline reconstructed segment time is shown. 
As can be seen, the time resolution with respect to segments of the Phase-2 system muon trigger primitives is of the order of a few ns, while the legacy system only gives trigger output in steps of bunch crossings (25~ns).

\section{Summary}
The Analytical Method for trigger primitive generation in Phase-2 for CMS DT chambers was here described. 
Results obtained by running a software emulator on Phase-2 simulations show good performance in terms of efficiency, time and spatial resolution, and rates. The new capability of around 1~ns digitisation in the TP generation is determinant for achieving these results.
Efficiencies for triggering at the right BX are uniform over the whole detector at values around 96-98$\%$ for prompt muons with p$_T>20$ GeV, at the challenging conditions of 200 pile-up events expected in HL-LHC. 
When considering ageing effects for the DT chambers, a decrease in efficiency can be observed in particular in the MB1 external wheels. 
However, this drop in efficiency does not significantly affect the performance of the L1 CMS trigger system.
Resolutions of the TP time, position, and slope are comparable to current results for offline segment reconstruction.
In particular, the bunch crossing time is obtained with about 2 ns resolution, and the difference in global position $\phi$ and direction $\phi_B$ for trigger primitives with respect to offline reconstructed segments are $<$ 50~$\mu$rad and $<$ 1~mrad respectively.
TP rates obtained from simulation are well within hardware constraints for Phase-2 bandwidth, specifically in the MB1 stations of the external wheels where the rates are larger.
The Analytical Method was implemented in firmware in a Virtex 7 FPGA. 
The current firmware is fully functional and capable of operating integrated into the CMS Trigger and DAQ system.
Firmware to Emulator  comparisons performed on a dedicated test bench using as input CMS collision data show already a good agreement for this early stage of implementation, with values of all relevant parameters agreeing at the level of the least significant bit for more than  97$\%$ of the primitives generated by the emulator.
During the Long Shutdown 2 of the LHC, one CMS DT sector (Wheel +2 Sector 12) was instrumented with the Phase-2 DT electronics prototype boards and integrated within the CMS DAQ system. This allowed an evaluation of the electronics and firmware implementation in a sustained campaign of cosmics-muon data taking,  already demonstrating  the large improvement in time resolution that can be achieved for Phase-2 TPs with respect to the legacy system.  The Phase-2 Slice Test is commissioned and will run when LHC resumes collisions in 2022.

\section{Conclusions}
The AM algorithm that was developed to reconstruct muon track segments on the CMS DT detector for HL-LHC has been validated with the worst expected conditions, showing excellent performance. The first hardware implementation has also been tested with cosmic data at CMS, confirming the good results. It is therefore expected a significant improvement in performance of the CMS DT trigger during HL-LHC. In the next years, efforts for porting this algorithm to the final FPGAs platforms and integration with the rest of CMS will proceed.

\section*{Acknowledgment}

We gratefully acknowledge the enduring support provided by the following funding agencies: the Bundesministerium f\"ur Bildung und Forschung, Germany; the National Research, Development and Innovation Fund, Hungary; the Istituto Nazionale di Fisica Nucleare, Italy; Agencia Estatal de Investigaci{\'o}n del Ministerio de Ciencia e Innovaci{\'o}n, Programa Estatal de Fomento de la Investigaci{\'o}n Cient{\'i}fica y T{\'e}cnica de Excelencia Mar\'{\i}a de Maeztu, and Plan de Ciencia, Tecnolog{\'i}a e Innovaci{\'o}n de Asturias, Spain.

\end{document}